\documentclass[]{aa}

\usepackage{lscape}
\usepackage{graphicx}
\usepackage{epsfig}
\usepackage{natbib}
\usepackage{longtable}
\usepackage{supertabular}
\usepackage{booktabs}
\bibpunct{(}{)}{;}{a}{}{,}
\usepackage{txfonts}
\begin{document}

   \title{Fe K${\alpha}$ line emission from the Arches cluster region -\\ evidence for ongoing particle bombardment?}
   \titlerunning{Fe K${\alpha}$ line emission from the Arches cluster region}
   \authorrunning{R. Capelli et al.}

   \author{R. Capelli\inst{1}, R.S. Warwick\inst{2}, D. Porquet\inst{3}, S. Gillessen\inst{1} and P. Predehl\inst{1}}

   \offprints{R. Capelli,\\ e-mail: capelli@mpe.mpg.de}

   \institute{Max-Planck-Institut f\"ur Extraterrestrische Physik, 
              Giessenbachstrasse 1, 85748 Garching, Germany
\and
	      Department of Physics and Astronomy, University of 
	      Leicester, Leicester LE1 7RH, UK
\and
	      Observatoire Astronomique de Strasbourg, Universit\'e de Strasbourg, CNRS, UMR 7550, 11 rue de l'Universit\'e, 67000 Strasbourg, France
}

   \date{Received ...; ...}

\abstract {Bright Fe-K$_{\alpha}$ line emission at 6.4-keV is a unique
characteristic of some of the dense molecular complexes present in the
Galactic  Center  region.  Whether  this  X-ray  fluorescence  is  due  
to the irradiation  of the  clouds by  X-ray photons  or is,
at least in part,  the result  of  cosmic-ray particle  bombardment,
remains  an interesting open question.}
{We  present the results of eight years of XMM-Newton observations of 
the region
surrounding the Arches  cluster in the Galactic  Center. We study
the spatial distribution and temporal behaviour of the Fe-K$_{\alpha}$
emission with  the objective of  identifying the likely source  of the
excitation.}
{We construct an Fe-K$_{\alpha}$  fluence map in
a narrow energy band of width 128~eV centered on 6.4-keV. We use this
to localize the brightest fluorescence features in the vicinity of the
Arches cluster.
We  investigate the variability of  the 6.4-keV
line emission of several clouds  through spectral fitting of  the 
EPIC~MOS data with  the use of a modelled background,  which avoids many of
the  systematics inherent  in local  background subtraction.   We also
employ spectral  stacking of both EPIC~PN and MOS data  to search for
evidence of  an Fe-K  edge feature imprinted  on the  underlying X-ray
continuum.}
{The lightcurves  of the  Fe-K$_{\alpha}$  line emission
from three bright molecular knots close  to the Arches cluster are found
to be constant over the  8-year observation window.  West of
the  cluster, however, we  found  a bright  cloud  exhibiting  the fastest
Fe-K$_{\alpha}$ variability yet seen in a molecular cloud in the Galactic
Center region. The time-averaged spectra of the molecular clouds reveal no
convincing  evidence of  the 7.1-keV edge feature,  albeit with  only weak
constraints. The EW of the 6.4-keV line emitted by the clouds near
the cluster is found to be $\sim~1.0$ keV.}
{The observed Fe-K$_{\alpha}$ line flux  and the high EW
suggest the fluorescence has a photoionization  origin, although excitation
by cosmic-ray particles is not specifically excluded.
For the three clouds nearest to the cluster, an identification of the source of 
photo-ionizing photons with an earlier outburst of Sgr~A* is however at 
best tentative.
The hardness of  the nonthermal  component associated with the
6.4-keV  line  emission  might be  best explained  in  terms  of 
bombardment by cosmic-ray particles from the Arches cluster itself. The 
relatively  short-timescale variability
seen in  the 6.4-keV line emission from  the cloud to the  West of the
cluster is most  likely the  result of illumination  by a
nearby transient X-ray source.}



   \maketitle

\section{Introduction}

The  Galactic  Center (GC)  is  a  bright  source of  diffuse  6.4-keV
fluorescent line  emission corresponding to the K$_{\alpha}$ transition in
neutral iron (Fe)  atoms (or  low-ionization Fe ions).  The spatial  
distribution of the line emission  is clumpy, but
at the same  time  widespread  across the whole region,  and  shows  a  good
correlation     with     that     of    molecular     clouds     (MCs)
\citep[][]{2007ApJ...656..847Y}. The three regions which have the most
prominent      6.4-keV     line      emission      are     Sgr      B2
\citep[][]{2009PASJ...61S.241I}, Sgr C \citep[][]{2009PASJ...61S.233N}
and  the  molecular  filaments  between  Sgr  A*  and  the  Radio  Arc
\citep[][]{2009PASJ...61S.255K}.  However, more  than a decade after
the       discovery      of       this       fluorescent      emission
\citep[][]{1996PASJ...48..249K},  the   mechanism  of  its  excitation
remains  a  puzzle.  The  fluorescence   may  be  the  result  of  the
irradiation of cold gaseous matter by hard
X-rays  photons with  energies above  7.1 keV  (the K-edge  of neutral
iron). Alternatively, the excitation might be  through the bombardment
of the gas by cosmic-ray  (CR) particles with energies above this same
threshold. In both cases the removal of a K-shell electron, is rapidly
followed  by   recombination  and   the  emission  of   a  fluorescent
K$_{\alpha}$  photon. In  X-ray  binaries and  active galactic  nuclei
(AGNs),  the observed  6.4-keV  line is  generally interpreted  as the
reprocessing of X-rays from the  central X-ray continuum source at the
surface of surrounding dense media, such as an accretion disc and/or a
molecular torus
\citep[e.g.,][]{1989MNRAS.238..729F,1997ApJ...477..602N}.  However, in
the case of  the fluorescencing structures seen in the GC  region, there is no
obvious persistent source bright enough to explain  the observed line fluxes. 
One  possible solution to this paradox is  that the X-ray illuminating source,  
from our offset perspective,  is highly  obscured.   Alternatively, such  a source  may
possibly have  been bright  in the past, but  is currently
faint.

\begin{figure*}[!Ht]
\begin{center}
\includegraphics[width=0.9\textwidth]{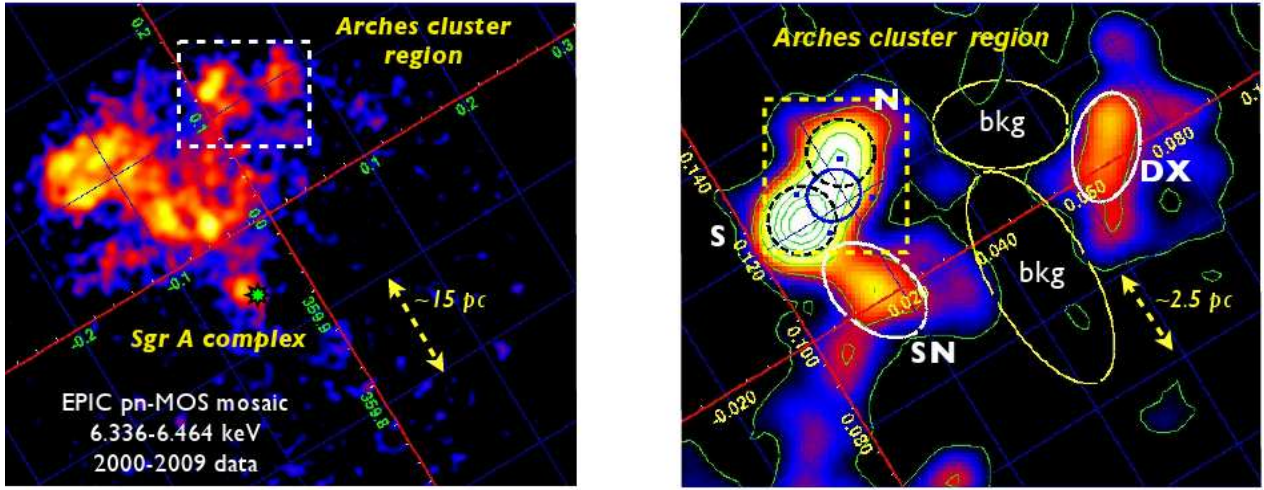}
\end{center}
\caption{$\textit{Left panel}$: Fe-K$_{\alpha}$ emission line map of the GC region. The data have been background subtracted and vignetting corrected. The position of Sgr A* is marked by a green star. The dashed white ellipse shows the AC region. $\textit{Right panel}$: A zoom into the AC region. The position of the AC 
is indicated by the blue ellipse at l$\sim$0.12$^{\circ}$, b$\sim$0.02$^{\circ}$. Bright 6.4-keV knots are highlighted by
dashed black circles (labelled N and S) and white ellipses (labelled SN at b$\sim$0.015$^{\circ}$ and 
DX at b$\sim$0.07$^{\circ}$). The region used for background accumulation is shown by the two yellow
ellipses (bkg). The dashed yellow box shows the sky region studied in \citet[][Fig.14]{2006MNRAS.371...38W}. 
The galactic coordinate grid is shown in red.} 
\label{imaging}
\end{figure*}

The   first   observations   of   6.4-keV   emission   from   the   GC
\citep[][]{1996PASJ...48..249K}  prompted   the  suggestion  that  the
required X-ray illumination might be associated with the past activity
of    Sgr     A*    \citep{1998MNRAS.297.1279S}.    In    particular,
\citet{2000ApJ...534..283M} suggested,  based on their  X-ray study of
the molecular cloud with the brightest 6.4-keV emission, Sgr B2, that 
Sgr  A* had  been in  an active phase  roughly three  hundred years
earlier.   This interpretation  has remained  in favour;  for example,
\citet{2010ApJ...719..143T}  have recently measured  the decay  of the
lightcurve of the scattered X-ray  continuum emanating from Sgr B2 and
argue that this  requires the period of intense activity  of Sgr A* to
have  ended between  75  and 155  years  ago. In  this  setting a  key
observational  fact is  that  Sgr A*  is  currently some six orders  of
magnitude fainter than required \citep[][]{2001Natur.413...45B}.
Even the brightest X-ray flares detected so far \citep[amplitude 100-160;][]{2003A&A...407L..17P,2008A&A...488..549P} are still 4 orders
of magnitude fainter than required and have too short a duration ($\sim$1 hour).
The  X-ray photon  irradiation scenario  is often  referred to  as the
X-ray Reflection  Nebulae (XRN) model, irrespective of  whether Sgr A*
is the actual source of the X-ray photon illumination.  

An alternative to the  XRN model is  to invoke the  incidence of CR  electrons and/or
protons as the excitation source \citep[e.g.  ][]{2003AN....324...73P,
2007ApJ...656..847Y,    2009PASJ...61..901D}.    Although   low-energy
(E$\lesssim$100  keV) CR  electrons are  favoured  over highly
energetic particles \citep[][]{2002ApJ...568L.121Y},  we note that the
cross-section for the production of inner shell ionization of Fe atoms
by  ultra-relativistic (Lorenz factor  $\gtrsim$100) electrons  is not
negligible;   in  fact,   it  amounts   to  about   1\%  of   the  Fe-K
photoionization  cross-section \citep[see][]{2003EAS.....7...79T}.  In
the case of  protons, the cross-section for the  ionization process is
highest for subrelativistic particles \citep[][]{2009PASJ...61..901D}.

One main way  to distinguish between the two  proposed scenarios is to
investigate the temporal variability  of the 6.4-keV line emitted from
the GC MCs, since fast variability  (typically on a timescale of a few
years) in  which reflection echos  appear to propagate at  roughly the
speed of light,  favours illumination by X-ray photons  rather than by
non-relativistic CR particles.  \citet[][]{2010ApJ...714..732P} argued
that the XRN model coupled  with the hypothesised past activity of Sgr
A*,  provided a  consistent  explanation for  the  complex pattern  of
variability seen in the molecular filaments located between Sgr A* and
the Radio Arc. However, it  remains unclear whether or not CR particle
bombardment  also  plays  a  role  in  the  excitation  of  the  X-ray
fluorescence from at  least some of these peculiar  MCs.  For example,
the  non-zero level  of the  Fe-K$_{\alpha}$  lightcurve prior  to the
onset  of an  episode of  enhanced emission  in the  molecular complex
known as the  ``Bridge'' \citep[][]{2010ApJ...714..732P}, could be due
to an underlying CR-induced line flux.

Here, we  report an X-ray study of  the MCs in the  surroundings of the
Arches cluster  (AC) based on observations carried  out by XMM-Newton.
This  region was not  studied in  detailed in  the recent  Chandra and
XMM-Newton     compilations     of     6.4-keV    cloud     properties
\citep[][]{2007ApJ...656L..69M,2010ApJ...714..732P},  although earlier
Chandra    results    for    the    AC   have    been    reported    by
\citet[][]{2002ApJ...570..665Y},  \citet[][]{2004ApJ...611..858L}, and
\citet[][]{2006MNRAS.371...38W}

\onltab{1}
{
\begin{table*}
\caption{Specifications for the selected OBSIDs. } 
\label{log_table}
\centering
\begin{tabular}{|c|c|ccc|}
\hline
OBSID & Obs Date & PN & MOS1 & MOS2 \\
 & yyyy-mm-dd &  cut/GTI/exp & cut/GTI/exp & cut/GTI/exp \\
\hline
\hline
0111350101 & 2002-02-26 & 0.8/38.590/40.030 & 0.5/42.262/52.105 & 0.5/41.700/52.120 \\
0202670501 & 2004-03-28 & 2.0/13.320/101.170 & 1.0/33.070/107.784 & 1.0/30.049/108.572 \\
0202670601 & 2004-03-30 & 2.0/25.680/112.204 & 1.0/32.841/120.863 & 1.0/35.390/122.521 \\
0202670701 & 2004-08-31 & 1.0/59.400/127.470 & 0.5/80.640/132.469 & 0.5/84.180/132.502 \\
0202670801 & 2004-09-02 & 1.0/69.360/130.951 & 0.5/94.774/132.997 & 0.5/98.757/133.036 \\
0402430301 & 2007-04-01 & 1.5/61.465/101.319 & 0.8/61.002/93.947 & 0.8/62.987/94.022 \\
0402430401 & 2007-04-03 & 1.5/48.862/93.594 & 0.8/40.372/97.566 & 0.8/41.317/96.461 \\
0402430701 & 2007-03-30 & 1.5/32.337/32.338 & 0.8/26.720/33.912 & 0.8/27.685/33.917 \\
0505670101 & 2008-03-23 & 1.25/74.216/96.601 & 0.5/73.662/97.787 & 0.5/74.027/97.787 \\
0554750401 & 2009-04-01 & 1.0/30.114/38.034 & 0.5/32.567/39.614 & 0.5/33.802/39.619  \\
0554750501 & 2009-04-03 & 1.0/36.374/42.434 & 0.5/41.376/44.016 & 0.5/41.318/44.018 \\
0554750601 & 2009-04-05 & 1.0/28.697/32.837 & 0.5/37.076/38.816 & 0.5/36.840/38.818\\
\hline
\end{tabular}
\tablefoot{In each of the instrument
related columns we report the threshold used for the Good Time Interval (GTI)
selection in the 10-12 keV lightcurve (in units of counts/s), the total GTI
exposure and the nominal duration of each observation.}
\end{table*}
}

Throughout this  paper, we  assume the
distance to the GC to be 8 kpc \citep[][]{2009ApJ...692.1075G}.
In Section 2 of this  paper we present the XMM-Newton observations and
describe  the data reduction  techniques we  have employed.   In Section 3, we
investigate  the  spatial distribution  of  the  Fe-K$_{\alpha}$  line
emission in the vicinity of the  AC and select four bright regions for
further temporal and spectral analysis. The study of the Fe-K$_{\alpha}$  line variability has been performed using background modelling, while for the
spectral analysis of the time  averaged spectra we employed a standard
background subtraction technique. We then  go on to discuss
our  results in  the context  of the  two main  scenarios  proposed to
explain the Fe fluorescent-line emission, namely the XRN model and the
CR  particle  bombardment  model   (Section  4).  Finally,  Section  5
summarises  our main results.

\section{Observations and data reduction}

We selected archival XMM-Newton data of the GC region largely targeted
at Sgr A*.   We reprocessed the data from both the  PN and MOS cameras
(\citet{2001A&A...365L..18S},  \citet{2001A&A...365L..27T})  with  the
tasks EPPROC and EMPROC in  the Science Analysis Software version 9.0.
In  order  to  mitigate   the  effects  of  occasional  high  particle
background rates in  the detectors, induced by bursts  of soft protons
incident on the satellite, we  constructed the 10-12 keV lightcurve of
the data from the whole field of view (FOV).  A threshold was then set
so as  to cut all the  peaks in the background  rate, thereby defining
the  Good  Time  Intervals  (GTIs)  for  the  time  filtering  of  the
observation.   Since  the  selected  observations  were  performed  in
varying conditions of  internal/particle background and orbital phase,
we  investigated  all the  lightcurves  independently  and selected  a
threshold   count   rate   separately  for   each   observation/camera
combination \citep[see also][]{2011A&A...525L...2C}.  The specifics of
the XMM-Newton observations employed  in the present work are reported
in  Table  \ref{log_table}.   Throughout  our analysis  we  have  only
selected single and double events (PATTERN$\le$4) for the PN and up to
quadruple events  (PATTERN$\le$12) for the MOS1 and MOS2  cameras. For
all the instruments a further screening involved selecting only events
marked as real X-rays (FLAG==0).

\section{Analysis and Results}

\subsection{The spatial distribution of the 6.4-keV line and its association 
with MCs}

A first task was to construct a fluence (integrated  flux over time) map  for the 
Fe-K$_{\alpha}$ line in  a narrow spectral window from which we might hope to 
identify and locate the brightest fluorescence regions.  We merged
the  GTI-filtered  event files for  the three EPIC  cameras (where
available) for each observation, using the SAS  task EMOSAIC.   
In  building the  image  of  the Fe-K$_{\alpha}$
emission,  we  assumed a  value  of  E/$\Delta$E=50 at 6.4 keV  for the  spectral
resolution (FWHM) of both the PN  and MOS1/2 cameras, corresponding to
a bandpass for the Fe fluorescence  signal of 6336-6464 eV. The resulting map of
the   6.4-keV  line   emission  is   shown  in the left panel of
Fig.\ref{imaging}; the AC is located at the edge of the XMM-Newton FOV  at 
the position R.A.=17$^{h}$45$^{m}$40.045$^{s}$,  DEC.=-29$^{\circ}$0'27.9''.
The general distribution of the Fe fluorescent emission evident 
in this figure is very  similar to that previously reported from both XMM-Newton
\citep[][]{2003AN....324...73P}    and Suzaku
\citep[][]{2009PASJ...61S.255K} observations.

\citet[][]{2003ApJ...590L.103Y} studied the non-thermal emission in the locality
of the AC and showed that the giant non-thermal filament of the Radio Arc 
(l$\sim$0.2$^{\circ}$) runs to the East of the AC with smaller thermal filaments 
lying to the north. The non-thermal emission originates in synchrotron radiation 
emitted by relativistic electrons in the strong magnetic fields within the 
Radio Arc structure.
Examination of a broadband (2--10 keV) version of Fig.\ref{imaging} 
shows that the general profile and extent of the AC inferred from the XMM-Newton
data is in good agreement with previous Chandra measurements
\citep[][]{2002ApJ...570..665Y, 2004ApJ...611..858L, 2006MNRAS.371...38W}.
The AC morphology consists  of  a  northern  core  component
\citep[designated as A1 and A2 in][]{2002ApJ...570..665Y} and a southern extended
component \citep[A3 in][]{2002ApJ...570..665Y}.  XMM-Newton separates these
north and south continuum components but is unable to resolve structures within them.  
The right panel of Fig.\ref{imaging} shows that there are three regions relatively close to the AC
which can be identified as distinct sources of 6.4-keV fluorescence. Two of these regions
coincide with the north (N) and south (S) continuum components defined above.
The third is located to the South-West of the S component (labelled as region SN).  
Moreover, the XMM-Newton observations also reveal a 6.4-keV bright knot 
further to the West of the N component (labeled as region DX).

There are some notable differences between the Fe-K$_{\alpha}$  line emission  map 
obtained  with XMM-Newton (Fig.\ref{imaging}, right panel) and the  one measured    by    Chandra
\citep[][]{2006MNRAS.371...38W}. First, XMM-Newton detected a northern 6.4-keV feature (knot N)
close to the AC itself,  which was not seen in the equivalent Chandra line image. With the
higher statistics  and the larger  effective area of EPIC  cameras, we also
discovered a prominent tapered feature (knot SN) that seems to be
connected to the knot S region. 

\begin{figure}[!Ht]
\begin{center}
\includegraphics[width=0.4\textwidth]{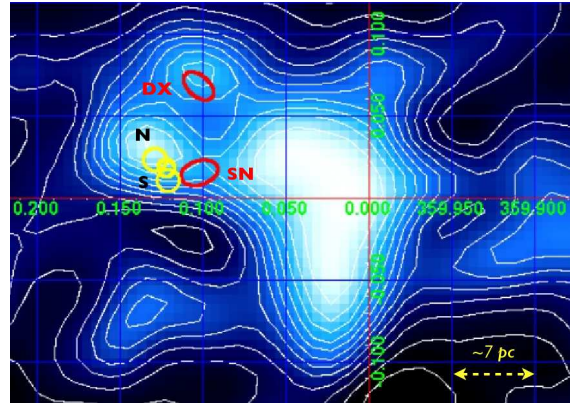}
\end{center}
\caption{Intensity map in galactic coordinates of the CS J=1-0 line 
\cite[]{1999ApJS..120....1T} integrated over the velocity range -40 km/s - 0 km/s.
The locations of the four bright knots identified in the XMM-Newton Fe-K$_{\alpha}$ image 
 are also indicated. }
\label{imaging_CS}
\end{figure}

To identify possible molecular  counterparts to  the  Fe-K$_{\alpha}$
bright knots  we used data from the  CS J=1-0 (48.99 GHz)  survey of the
Central Molecular Zone (CMZ) \citep[][]{1999ApJS..120....1T}.
The results are  shown in Fig.\ref{imaging_CS}, where the blue colour
scale and the contours represent  the CS J=1-0 line emission intensity
integrated  over the  -40  km/s to  0  km/s velocity  range. The
locations of the bright 6.4-keV knots are also indicated.
A CS enhancement at l$\sim$0.13$^{\circ}$, b$\sim$0.03$^{\circ}$ coincides
positionally with knot N. This molecular feature has a velocity of -30 km/s. 
Lower surface brightness  molecular structures are also seen close to the SN and DX knots.
We note that the CS emission at the four knot locations spans a
spread of velocities which, although  narrow enough to likely rule out the 
possibility that the molecular material is  widely distributed
along the line of sight, is sufficiently broad to suggest that the 
dynamics of the MCs in this region are dominated by turbulence.

\begin{table}[ht]
\caption{Physical properties of the 6.4-keV emitting clouds.}
\label{clouds}
\centering
\begin{tabular}{ccccc}
\hline
cloud & size (pc) & N$_{H}$(cm$^{-2}$) & \#H (10$^{59}$) & $\tau_{T}$ \\
\hline
N & 0.92 & 4$\times$10$^{22}$ & 9.6 & 0.0266 \\
S & 0.92 & 2$\times$10$^{22}$ & 4.8 & 0.0133 \\
SN & 1.6$\times$1.0 & 2$\times$10$^{22}$ & 9.0 & 0.0133 \\
DX & 1.5$\times$0.9 & 2$\times$10$^{22}$ & 7.6 & 0.0133\\
\hline
\end{tabular}
\tablefoot{Here, we report the cloud size in parsecs (in radius or semi-axis), the equivalent 
H column density through the cloud, the estimated total number of H atoms 
inside the cloud and the optical depth of the cloud to Thomson scattering.}
\end{table}

\begin{table*}[!Ht]
\caption{Fluxes of the Fe-K$_{\alpha}$ line.}
\label{ka_fluxes}
\centering
\begin{tabular}{|c|cccccc|cc|}
\hline
region & Feb02 & Mar04 & Sep04 & Apr07 & Mar08 & Apr09 & Wtd Mean & $\chi^{2}_{red}$\\ 
\hline
N & 3.4$\pm$1.7 & 4.0$\pm$1.0 & 4.8$\pm$1.0 & 3.4$\pm$1.0 & 3.4$\pm$1.1 & 4.4$\pm$1.1 & 3.9$\pm$0.4 & 0.4\\
S & 4.9$\pm$1.9 & 7.2$\pm$1.3 & 7.8$\pm$1.2 & 6.5$\pm$1.0 & 6.6$\pm$1.4 & 7.4$\pm$1.3 & 6.9$\pm$0.5 & 0.4\\
SN & 4.7$\pm$1.6 & 3.9$\pm$1.0 & 5.3$\pm$0.9 & 4.5$\pm$0.8 & 6.3$\pm$1.2 & 3.5$\pm$0.9 & 4.6$\pm$0.4 & 0.9\\
DX & 1.2$\pm$1.1 & 3.1$\pm$0.9 & 4.7$\pm$0.9 & 3.3$\pm$0.7 & 1.9$\pm$0.8 & 1.7$\pm$0.7 & 2.7$\pm$0.3 & 2.1\\
\hline
\end{tabular}
\tablefoot{The values reported in the Table are the weighted means of the MOS1\&2 measured fluxes, in units of 10$^{-6}$ photons/cm$^{2}$/s. The 
different columns refer to different datasets (see text). The Apr07, 
Mar08 and Apr09 fluxes for region D are based on  MOS 2 data only, as a
consequence of the damaged sustained by CCD6 in MOS 1 on 9 March 2005 \citep[][]{2006ESASP.604..943A}.}
\end{table*}

\begin{figure*}[!Ht]
\begin{center}
\includegraphics[width=0.95\textwidth]{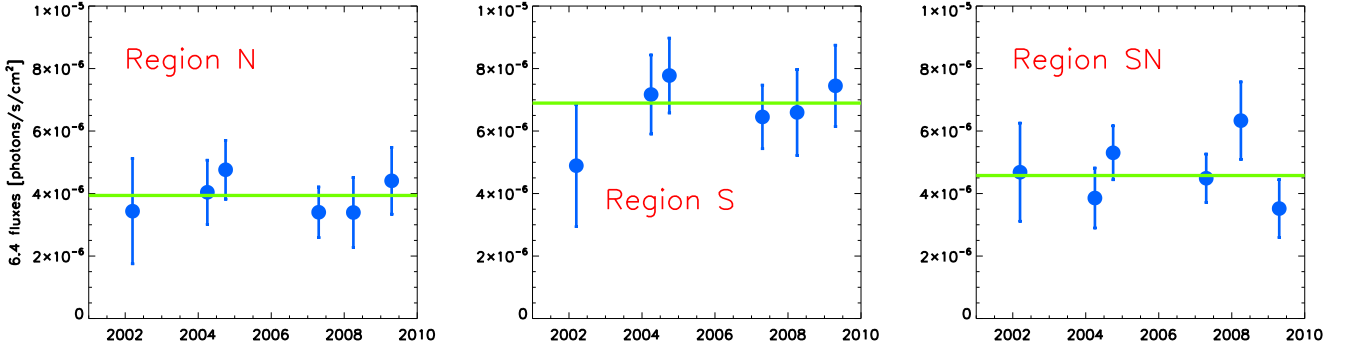}
\end{center}
\caption{Lightcurves of the Fe-K$_{\alpha}$ line flux from the MCs in the vicinity of the AC. 
The flux values are in units of 10$^{-6}$ photons/cm$^{2}$/s. The green horizontal
lines show the weighted means of the data points. 
}
\label{pmulti}
\end{figure*}

Hereafter, we take the view that the four 6.4-keV knots represent coherent molecular
structures in the vicinity of the AC.
On this basis, Table  \ref{clouds}  details  the  physical  parameters of
each MC.  The angular size of the  clouds have been estimated by
eye and then converted to a linear size.
For  the hydrogen column density through the N and S clouds,  N$_{H}$, 
we  use  the  values   reported  by \citet{2009ApJ...694..943A} which in 
turn were derived  from CS molecular  emission.  Because
these authors did not report the equivalent H column densities for the
other two knots, we assumed these to be the same as the
region S (since the emission  contours of  the CS  J=1-0 line,
at the locations of these knots are similar to those in the S region).
We caution that the estimate of the N$_{H}$ of a MC in
the GC  is strongly dependent  on the method  used to measure  it. For
example, for the large G0.13-0.13 molecular complex, 
two extreme values have  been inferred for N$_{H}$, namely 4$\times$10$^{22}$ cm$^{-2}$
\citep[][using CS  J=1-0 line]{2009ApJ...694..943A} and a much higher value of
10$^{24}$  cm$^{-2}$ \citep[][based on the H$^{13}$CO$^{+}$
J=1-0 line]{2006ApJ...636..261H}.
Finally, we calculated the  number of H atoms in the MCs  and the value of
the  optical depth  for Thomson  scattering ($\tau_{T}$).  
The estimates in Table \ref{clouds} will be used in Section 4 to investigate 
the origin of the Fe-K$_{\alpha}$ line emission.

\subsection{Fe-K$_{\alpha}$ line flux and variability - background modeling}

We are interested in determining both the absolute photon flux in the 
Fe-K$_{\alpha}$ line and in searching for temporal variability
in this signal. To this end, we extracted the spectra for each of the four 
bright 6.4-keV knots for each observation and then, for observations separated
by only a few days, combined the data to give a sampling at six epochs as follows:
February 2002 (1 observation), March 2004 (2 observations), August/September 2004 
(2 observations), March/April 2007 (3 observations), March 2008 (1 observation), 
and April 2009 (3 observations).

The 6.4-keV emitting knots are not visible in the  full  band (2--10 keV)  
X-ray image, since their continuum signal is hard to distinguish against the
strong  diffuse  X-ray emission which  permeates the  GC region.  
Indeed, the  main contribution
to the  2--10 keV  spectrum of these extended sources is the thermal 
emission from the GC region. The
best choice in performing a self-consistent spectral analysis of these
low  surface brightness  features is  to model  the background as a component within 
the spectral fitting rather than subtracting it prior to the fitting.
This technique avoids  many systematic errors due  to the
contamination by foreground emission 
\citep[][]{2004A&A...419..837D,2008A&A...486..359L}, but has the disadvantage that we
are forced to exclude PN data (since at present, no detailed study of the 
instrumental background for the PN EPIC-camera has  been provided). 

The  model used in the spectral fitting accounts for:

\begin{itemize}
\item{Photoelectric absorption \citep[WABS,][]{1983ApJ...270..119M}. 
All the emission components are subject to soft X-ray absorption so as
to account for the high column density along the line of sight to the GC region.}
\item{Galactic thermal diffuse emission \citep[2 $\times$ APEC,][]{2001ApJ...556L..91S}. 
Two thermal components  are needed to explain the spectral 
shape of the GC thermal emission: one $\textit{warm}$ with a temperature 
of $\approx$ 1 keV \citep[][]{2009PASJ...61..751R} and one $\textit{hot}$ with a temperature in the 
range 5-7 keV \citep[][]{2007PASJ...59S.245K}. The higher temperature plasma is needed to
constrain the fit in the region of Fe-line complex.
The metallicities of the two thermal plasmas have been fixed to twice
solar. \cite{2002A&A...382.1052T} showed that the equivalent 
width of the S and Fe lines in the spectrum of the Galactic 
Ridge decrease as one moves away from the GC,  with higher than solar
metallicities required for elements such as Si, S and Fe in the 
diffuse plasma permeating the GC region. 
Recently, \citet[][]{2010PASJ...62..423N} also measured a 
supersolar metallicity for the hot plasma in the GC region.
The twice solar constraint 
results in a broadly satisfactory fit to all our spectra, 
whereas use of a solar metallicity produces high residuals in all 
the soft X-ray (i.e., $\lesssim$ 3-4 keV) emission lines.} 
\item{The Cosmic X-ray Background (POWER-LAW). The spectral shape 
of this component is modelled as a power-law continuum with a photon index 
fixed at $\Gamma$=1.4 \cite[][]{2007ApJ...671.1523H}. The normalization of 
this component was set to 10.9 photon cm$^{-2}$ s$^{-1}$ sterad$^{-1}$ 
keV$^{-1}$ \citep{2007ApJ...671.1523H}.}
\item{A non-thermal hard X-ray component (POWER-LAW). The non-thermal 
continuum emission directly linked to the Fe-line fluorescence. 
We fixed the slope of the  power-law component associated with the source
of  fluorescence to  $\Gamma$=1 in  order to  give consistency  in the
measurement    of   the    6.4-keV   flux    across    the   different
datasets\footnote{The value of $\Gamma$  was chosen after studying the
stacked spectrum of region S (see \S3.3), which shows the most intense
6.4-keV  line emission, and  therefore has  the best  statistics. This
analysis gave  $\Gamma$ = 1.0$^{+0.1}_{-0.2}$  consistent with earlier
estimates  from   Chandra  ($\Gamma$=1.3$^{+1.4}_{-1.1}$)  and  Suzaku
($\Gamma$=0.72$^{+0.68}_{-0.72}$)
\citep[respectively,][]{2006MNRAS.371...38W, 2007PASJ...59S.229T}.}.
}
\item{Fe fluorescent lines (2 GAUSS). The two spectral lines with gaussian profiles
were used to represent 
the Fe-K$_{\alpha}$ and Fe-K$_{\beta}$ fluorescent lines at
6.4 keV and 7.05 keV. The flux of the Fe-K$_{\beta}$ line was fixed at
0.11 times that of the K$_{\alpha}$ component \citep[][]{2009PASJ...61S.255K}.}
\item{The particle background (POWER-LAW). We used the values reported
in  \citet{2008A&A...486..359L}   for  the  slopes   of  the  particle
component in the MOS1 and MOS2 spectra, respectively $\Gamma$=0.24 and
$\Gamma$=0.23.  Since  this instrumental background is  not focused in
the detector, it does not need to be convolved with the instrument
response.}
\end{itemize}

The  statistics  of  each  spectrum  (per region,  per  epoch)  differ
markedly from  one another. Since the construction  of the lightcurves
of the 6.4-keV line flux must  be done in a consistent way, we decided
to fix the temperature of the  hot plasma component at 6.5 keV \citep[][]{2007PASJ...59S.245K},
given  that  variations  in  the temperature  of  this
component  have the  potential  to unduly  influence  the measured  Fe
K$_{\alpha}$ line flux.  Note, however, that the temperature of the warm
thermal emission was taken  to be a free parameter, as  were the 
normalizations of the two thermal plasma components.
The warm component temperature ranges inferred for the  N, S and SN
clouds are 0.5-2 keV, 0.7-1.3 keV and $\sim$ 1 keV respectively, values
which are  in good agreement with  those obtained in  Section 3.3 from
the analysis of the stacked spectra. These temperatures are marginally
lower  than the value  of 1.7  keV measured  for the  core of  the AC,
\citep[][]{2011A&A...525L...2C},  hinting at  a  decrease in  the warm  plasma
temperature as  one moves  away from the  cluster until  an "ambient"
medium with kT $\approx$ 1 keV is reached.  Because of the reduced signal
to noise ratio, all the spectra of the DX  region were fitted with
the warm  plasma temperature fixed at the  ambient value, (i.e.
kT=1 keV).

Because of the  different slopes of the particle  component in the MOS
cameras,  the   MOS1  and  MOS2   spectra  were kept separate, giving 
two  spectral datasets for
each of  the four regions (N, S, SN, DX)  for each of the  six epochs.
Given the restricted photon statistics, we used the  Cash
statistic  rather than  the  $\chi^{2}$.
Prior to spectral fitting we  grouped  the channels in  each spectrum  
with the  GRPPHA tool in  order to  have a minimum  of 1  
count/bin  \citep[][]{2004A&A...419..837D}.  
The   photon fluxes in the 6.4-keV line determined from this analysis
are reported in Table \ref{ka_fluxes}.
In the last two columns of this table we also report the weighted mean across the 
six measurement epochs and the reduced $\chi^{2}$ for the constant
flux hypothesis.
Fig.\ref{pmulti} shows the lightcurves of the 6.4-keV line for the three 
MCs in the immediate vicinity of the AC, together with the weighted
mean flux. We find that the 6.4-keV line  fluxes from all of these knots
(regions N, S and SN) are constant with time.

\begin{figure}[ht]
\begin{center}
\includegraphics[width=0.4\textwidth]{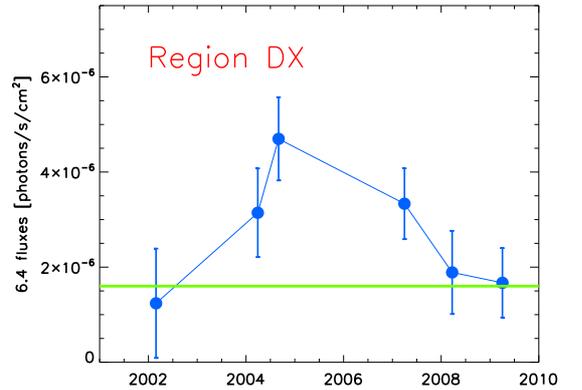}
\end{center}
\caption{Lightcurve of the 6.4-keV line for the DX cloud. 
The weighted mean (1.6$\pm$0.5$\times$10$^{-6}$photons/cm$^{2}$/s) 
of the first and last data points is shown as the horizontal green line.
The emission reaches a maximum in September 2004 with a 
line flux roughly three times the base level. 
}
\label{lc_destra}
\end{figure}

The DX cloud  is $\sim$ 3 arcmin (about 7 pc  in projection) to the West
of the  AC, in a region which appears to be disconnected from the  
cluster itself  and from the  other 6.4-keV bright knots   studied   
in   this   paper.  The  lightcurve  of  the  Fe
K$_{\alpha}$ emission from the DX cloud shows clear evidence for time variability 
(Fig.\ref{lc_destra}), which is supported by the high value of
$\chi^{2}_{red}$ for these data with respect to the constant flux 
hypothesis (Table \ref{ka_fluxes}). 
Clearly, the temporal behavior 
of this particular cloud is very different to that of the three bright knots
closer (in projection) to the AC. Its temporal behaviour is also
different to that reported for all the other MCs in the GC
\citep[][]{2009PASJ...61S.241I, 2010ApJ...714..732P}. In the DX knot
we clearly see both an increase and decrease in the  Fe fluorescent
flux rather than solely an increasing (or decreasing) trend over
the monitoring period. The Fe-K$_{\alpha}$ flux from this cloud appears to
have increased by a factor of three over a period of about 2.5 years 
with a peak in September 2004 after which, over a timescale of a few years,
it faded back to its pre-outburst level. Interestingly the average of the 
first and last points in the lightcurve is not zero, suggesting
that the outburst was superimposed on a base level. 
The variability exhibited by this cloud is the $\textit{fastest}$ yet reported
for the GC region.

\subsection{Analysis of the time averaged spectra - stacking + background subtraction} 

We  have also carried out an analysis  of the stacked PN and MOS
spectra of the four 6.4-keV knots with the objective of determining
the equivalent width (EW) of the Fe-K$_{\alpha}$ fluorescent line with respect
to the associated continuum. A second objective was to check for an Fe-K absorption 
edge at 7.1 keV imprinted on the same continuum \citep[][]{1998MNRAS.297.1279S, 2000ApJ...534..283M}. 
Because of the stacking of the  data (i.e. the addition of
all the spectra across the set of observations), the photon 
statistics were much improved,  which  allowed  us to employ a more
standard background subtraction technique and therefore include the EPIC-PN data 
in the study. We again used the C-statistics because of the low number of counts per bin in the spectra; indeed, when fitting with the $\chi^{2}$ statistics the $\chi^{2}_{red}$ values are always too small (about 0.2-0.3) to be confident with the best fit results. Moreover, for what concerns the spectral analysis of the MCs in the vicinity of the AC, we notice that the C-statistics give always tighter constraints on the spectral parameters.

\begin{table}[ht]
\caption{Results for the spectral analysis of the stacked spectra.}
\label{number_sigma2}
\centering
\begin{tabular}{c|cccc}
\hline
 & N & S & SN & DX \\
\hline
N$_{H}$ & 9.5$\pm$1.5 & 10.1$\pm$0.7 & 8.5$^{+4.0}_{-3.4}$ & 6.0 (fixed) \\
kT (keV) & 1.8$\pm$0.3 &  1.6$\pm$0.1 & 1.0$^{+1.0}_{-0.5}$ & - \\
norm$_{kT}$ & 4.7$^{+2.4}_{-1.5}$ & 9.2$^{+2.5}_{-2.4}$ & 2.8$^{+17.7}_{-2.4}$ & - \\
F$_{6.4}$ & 3.2$\pm$0.5 & 6.2$\pm$0.6 & 3.2$\pm$1.0 & 1.9$\pm$0.5 \\
norm$_{POW}$ & 2.0$\pm$0.3 & 4.4$\pm$0.4 &  1.8$^{+0.4}_{-0.2}$ & 0.5$\pm$0.1 \\
$\tau$ & $\leq$ 0.4 & $\leq$ 0.6 & $\leq$ 1.8 & - \\
EW$_{6.4}$ & 1.0$\pm$0.4 & 0.9$\pm$0.2 & 1.1$\pm$0.4 & 2.6$^{+2.1}_{-1.1}$ \\
C-stat & 2337.60 & 2280.52 & 2335.05 & 453.92 \\
\hline
\end{tabular}
\tablefoot{Here we report the total column density (N$_{H}$, in 10$^{22}$ cm$^{-2}$)
inferred from the X-ray measurements, the temperature and the normalization of the APEC thermal plasma (in units of 10$^{-18}$$\int$n$_{e}$n$_{H}$dV/4$\pi$D$^{2}$, where n$_{e}$ and n$_{H}$ are the electron and H densities in cm$^{-3}$, and D the distance to the source in cm), the flux of the 
Fe K$_{\alpha}$ line (in units of 10$^{-6}$ photons/cm$^{-2}$/s), the  normalization of the powerlaw component (10$^{-5}$ photons/keV/cm$^{-2}$/s at 1 keV), the EW of the Fe-K$_{\alpha}$ line (in keV) with respect to the powerlaw component, the upper limit to the optical depth of the Fe-K edge, and the C-stat values for the best fit model (the degrees of freedom are 2119 for the regions N, S and SN, 397 for the region DX). }
\end{table}

The background spectra for the PN and MOS channels were built
by stacking  all  the  background  spectra collected  for  the  different
datasets. The region selected for  the background accumulation corresponds
to the two ellipses with semi-axes 0.8$\times$0.5 arcmin and 1.4$\times$0.6 arcmin
shown in the right panel of Fig.\ref{imaging}. The spectral  
model used to fit the data comprises a  collisional  ionized  \textit{warm}  plasma (APEC, with temperature and normalization free parameters)
and a power-law continuum with associated  Fe-K$_{\alpha}$ and Fe-K$_{\beta}$ lines
at 6.4  and 7.05  keV (K$_{\beta}$/K$_{\alpha}$=0.11). All  the emission components
were then subject to absorption in the ISM along the line of sight (WABS, common to both the APEC and the power law). For this
analysis the slope of  the non-thermal continuum  radiation associated with
the fluorescence was again fixed at $\Gamma$=1.
The use of a local background removes the requirement for two
of the spectral components included previously, namely the hot thermal 
component and the cosmic X-ray background, both of which can be assumed
to have constant surface brightness across the AC region.

\begin{figure*}[!Ht]
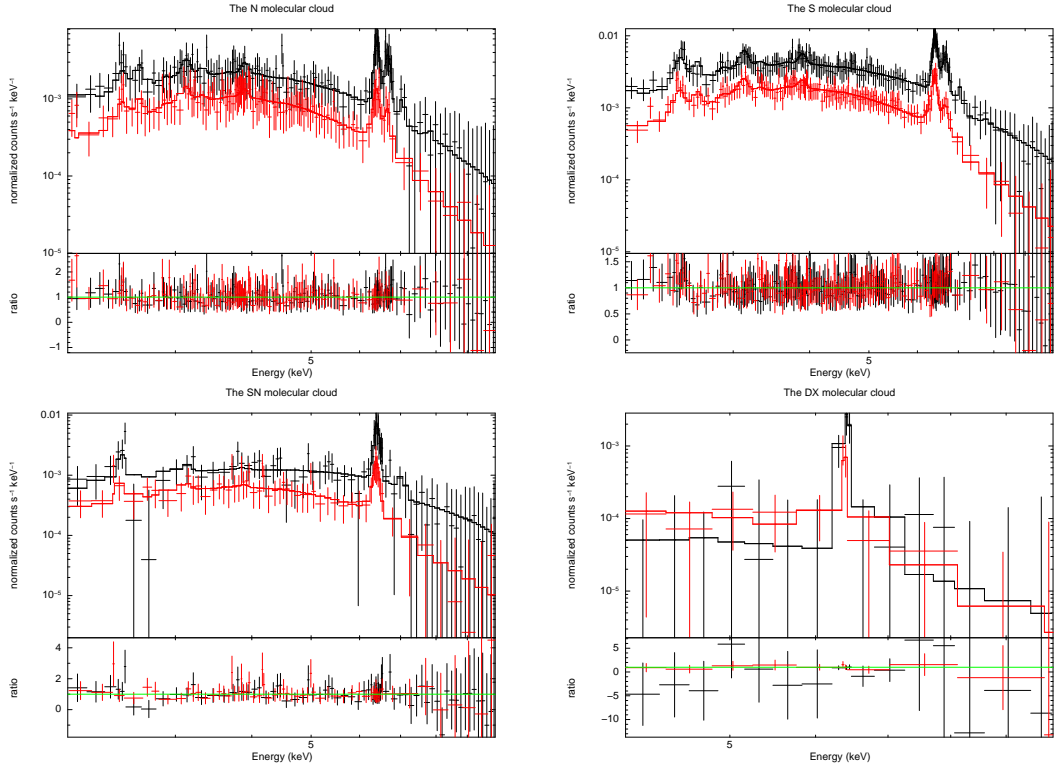

\begin{center}
\includegraphics[width=0.275\textwidth,angle=-90]{fig5a.eps}
\hspace{0.75cm}
\includegraphics[width=0.275\textwidth,angle=-90]{fig5b.eps} \\
\includegraphics[width=0.275\textwidth,angle=-90]{fig5c.eps}
\hspace{0.75cm}
\includegraphics[width=0.275\textwidth,angle=-90]{fig5d.eps}
\end{center}
\caption{PN (black) and MOS (red) stacked spectra of the four selected MCs in the
2-10 keV (4-10 keV for the DX cloud) energy range, together with the ratio between the data points and the best fit model. The top panels show the spectra of the  N and S molecular complexes,
with the bottom panels similarly presenting the spectra of SN and DX clouds.}
\label{pmulti_edge}
\end{figure*}

The results are summarised in  Fig.\ref{pmulti_edge} and
Table \ref{number_sigma2}\footnote{The values quoted here are based on the MOS data only. We do not consider the PN values here because of likely
systematic errors in the background subtraction. The fluxes of the
6.4-keV line in the background-subtracted PN spectra are generally
lower than those determined from the equivalent MOS spectra such that,
if the PN measurements are included, then the line fluxes based on the stacked spectra are not consistent with the MOS-only values derived in Section 3.2.
We tried also to fit the stacked PN spectra modeling the background with a simple powerlaw component ($\Gamma$=0), and got consistent results, both between PN and MOS cameras, and with the fluxes measured in Section 3.2. Therefore, we associate these systematics in PN measurements of the 6.4-keV emission to the background subtraction.}. 
Regions N and S have rather similar
spectra, in which contamination by residual thermal emission (not removed by the background
subtraction) is
clearly present in the form of the helium-like Fe line at  6.7 keV and 
other helium-like lines  of other elements  (S, Ar,  Ca) at  lower energies.
In these regions, close to the AC core, we derive a temperature of the \textit{warm} plasma consistent with that previously measured in the core \citep[][]{2006MNRAS.371...38W,2007PASJ...59S.229T,2011A&A...525L...2C}.
The X-ray spectra of  both complexes exhibit high line-of-sight
absorption and a  relatively high EW for the neutral iron
Fe K$_{\alpha}$  line  ($\sim$ 1.0  keV).
The high  value  of  the EW might be interpreted
as evidence for X-ray irradiation as the origin
of this fluorescence. For this reason, we also investigated
whether there is any evidence for the Fe-K absorption  edge, by applying
a multiplicative EDGE model (with a fixed energy of 7.1 keV)  to the 
power-law continuum component; however, on the basis of the spectral fitting
there was no requirement for such a feature. We could only measure an upper limit to the optical depth of this absorption edge (see Table \ref{number_sigma2}).
However, 
the detection of an optical depth of about 0.1-0.2 
(equivalent to N$_{H}\sim$10$^{23}$cm$^{-2}$) at the Fe-K absorption edge energy (7.1 keV)
is challenging when the  statistics are limited  and the  spectrum above 7  keV is
potentially contaminated by systematic errors in the subtraction of
the instrumental background
(which  has a  non-uniform pattern in the detector).

The  spectral parameters  inferred for  the SN  region are  similar to
those obtained for the N and S clouds, except for the marginally lower
temperature inferred for the warm plasma  (kT $\sim$ 1 keV) which, as noted
earlier, may plausibly  be a consequence of the increased distance from
the AC core.   A high value for the  EW of the 6.4 keV line  (EW $\sim$ 1.1
keV) is again  evident, with no requirement for  an absorption edge at
7.1 keV.

In the  case of the DX cloud  the derived spectrum, which  is shown in
the lower  right panel of Fig.5,  is very strongly  photon limited. In
fact the statistics are so poor  that no net signal was detected below
4  keV and  no warm  plasma component  was required  in the  fit.  The
inferred  2-10 keV  photon  flux  based on  the  spectral fitting  was
5.9$\pm$3.5$\times$10$^{-6}$ photons/cm$^{2}$/s, that is almost an order of magnitude lower
than the continuum fluxes inferred for the other knots. The measurement of an Fe-K absorption edge could not be done.

\subsection{Summary of the results}

The results of Section 3 can be summarized in five main points:

\begin{itemize}
 \item {Fe-K$_{\alpha}$ line variability (N-S-SN): the  lightcurve of the  6.4-keV line  flux has
been found  to be  constant in the  regions N,  S and SN.}
  \item{Fe-K$_{\alpha}$ line variability (DX): we measured a fast variability of the Fe fluorescent line  in the DX cloud, with a timescale of about 2-3 years.}
  \item{EW of  the 6.4-keV line:  All four MCs  studied have high values  ($\sim$ 1
keV)  for  EW of  the  Fe fluorescent  line,  although  the errors,
particularly for the DX cloud, are relatively large.
Unfortunately, the 90\%  confidence range for the EW is too wide to
place tight constraints on the excitation mechanism of Fe fluorescence. }
  \item{Fe-K edge at 7.1 keV: we failed to detect an absorption edge imprinted
on the power-law component in any  of the MCs - much better statistics
would be  required to properly constrain  the optical depth  of such a
feature.}
  \item{Hardness of the non-thermal emission: we measured a hard spectral slope for the non-thermal emission for the region S (i.e. $\Gamma$=1.0$^{+0.1}_{-0.2}$) with the  spectra  of  the  other clouds compatible with the
presence of such a component. }
\end{itemize}

\section{Discussion}

The CMZ contains approximately 10\% of the molecular gas of  the entire 
Galaxy largely in the form of dense clouds \citep[][]{1996ARA&A..34..645M}.
The bright Fe-K$_{\alpha}$ line emission at 6.4 keV  is a peculiar feature
of the MCs located within this region. The  6.4-keV line is produced by
the fluorescence of neutral (or near-neutral) Fe atoms subject
to irradiation by hard X-rays  (with energies above the Fe-K edge at
7.1  keV) and/or bombardment by CR  particles with  energies in the range
from 7.1 keV up to, potentially,  many 100's of keV. 
More than  a decade after  its first  detection, the combination of 
circumstances which give rise to the bright Fe-K$_{\alpha}$ emission in the
GC remains a matter of discussion   \citep[e.g.][]{2007ApJ...656..847Y,
2009PASJ...61S.255K,  2009PASJ...61..901D}.  In a recent  development, 
\citet[][]{2007ApJ...656L..69M} and \citet[][]{2010ApJ...714..732P} have studied 
the variability of the 6.4-keV line flux seen in several MCs 
lying between Sgr A* and the GC
Radio Arc. These authors interpret their results in terms of 
the photoionization of molecular material by hard X-rays  produced in 
a  past ($\sim$100  years  ago) bright  (L$_{2-10}\sim$10$^{39}$ erg/s) 
flare on the SMBH at the center of the Galaxy. 
On the other hand, some authors  are not fully convinced by
the Sgr A* outburst model 
\citep[e.g.,][]{2003AN....324...73P,2007ApJ...656..847Y,2009PASJ...61..901D}.

Here, we interpret the results of our study of four 6.4-keV
bright knots located within $\sim$ 3 arcmin of the AC. Projected onto the plane of the sky
the separation of these knots from the AC is no more than $\sim 7$ pc;
however, we have limited information on the relative
line-of-sight locations, so the association of these structures 
with the AC is only tentative.

\subsection{The XRN hypothesis}

The three knots closest (in projection) to the AC exhibit 
a constant Fe-K$_{\alpha}$ line flux and it is tempting
to imagine that the AC itself might be the source of the excitation, 
whether via photons or particles.  However, in the case of the former the
average X-ray luminosity of the AC is nowhere near sufficient
to produce the total amount  of fluorescence observed.
On the  other hand, if the AC entered a putative high activity state in which its
X-ray luminosity exceeded 10$^{37}$ erg/s, then it would be possible to
explain the observed 6.4-keV  fluxes of the nearby clouds in the context 
of the XRN model.
\citet[][]{2011A&A...525L...2C} recently reported X-ray flaring activity
within the AC, most likely originating from stellar winds interactions in
one or more massive binary systems,  but even at its peak the measured 
luminosity remains three order of magnitude lower than that cited above.  
An alternative is to consider the temporary brightening of
an X-ray binary source close to or even within the AC. Then the requirement
is for the bright state to have lasted for more than 8 years 
at or above $L_x \sim 10^{37}$erg/s; this is much  longer than the 
typical flaring timescales of  X-ray transient  sources
\citep[e.g.,][]{2010A&A...524A..69D}.

\subsubsection{Fe-K$_{\alpha}$ variability}

\begin{table*}[ht]
\caption{Cloud parameters and distances inferred in the XRN/Sgr A* outburst model.}
\label{clouds_distances}
\centering
\begin{tabular}{c|cccc|cccc|cccc|cccc}
\hline
cloud &  & N & & &  & S &  & & &  SN  &  & &  & DX & &  \\
\hline
\hline
z & 1z$_{\odot}$ & 1.25z$_{\odot}$ & 1.5z$_{\odot}$ & 2z$_{\odot}$ & 1z$_{\odot}$ & 1.25z$_{\odot}$ & 1.5z$_{\odot}$ & 2z$_{\odot}$ & 1z$_{\odot}$ & 1.25z$_{\odot}$ & 1.5z$_{\odot}$ & 2z$_{\odot}$ & 1z$_{\odot}$ & 1.25z$_{\odot}$ & 1.5z$_{\odot}$ & 2z$_{\odot}$ \\
R & 0.92 & 0.92 & 0.92 & 0.92 & 0.92 & 0.92 & 0.92 & 0.92 & 1.27 & 1.27 & 1.27 & 1.27 & 0.90 & 0.90 & 0.90 & 0.90 \\
F$_{6.4}$ & 3.9 & 3.9 & 3.9 & 3.9 & 6.9 & 6.9 & 6.9 & 6.9 & 4.6 & 4.6 & 4.6 & 4.6 & 4.7 & 4.7 & 4.7 & 4.7 \\
$\Omega$ & 7.6 & 6.1 & 5.1 & 3.8 &          26.8 & 21.4 & 17.9 & 13.4 &            17.9 & 14.3 & 11.9 & 8.9 &          18.0 & 14.4 & 12.0  & 9.0 \\

d & 52.8 & 58.9 & 64.4 & 74.6 &             28.1 & 31.4 & 34.4 & 39.7 &            47.5 & 53.1 & 58.2 & 67.3 &         33.5 & 37.5 & 41.1 & 47.4 \\

d$_{p}$ & 27.6 & 27.6 & 27.6 & 27.6 & 26.3 & 26.3 & 26.3 & 26.3 & 23.3 & 23.3 & 23.3 & 23.3 & 27.5 & 27.5 & 27.5 & 27.5 \\

d$_{l.o.s.}$ & 45.0 & 52.0 & 58.2 & 69.3 &          9.9 & 17.2 & 22.2 & 29.7 &                41.4 & 47.7 & 53.3 & 63.1 &              19.2 & 25.5 & 30.5 & 38.6 \\

\hline
\end{tabular}
\tablefoot{Here, Z is the metallicity 
assumed for the MC, R the radius of the cloud as seen from Sgr A* (pc), F$_{6.4}$ the flux measured in the Fe-K$_{\alpha}$ line 
(photons/cm$^{2}$/s), $\Omega$ the solid angle (scaled to the whole sky, in units of 10$^{-5}$) subtended by the MC 
from the perspective of Sgr A*, d(pc) the distance of the MC from Sgr A*, d$_{p}$(pc) and d$_{l.o.s.}$(pc) the distance
projected onto the plane of the sky and along the line of sight, respectively. }
\end{table*}

\begin{figure*}[ht]
\begin{center}
\includegraphics[width=0.4\textwidth]{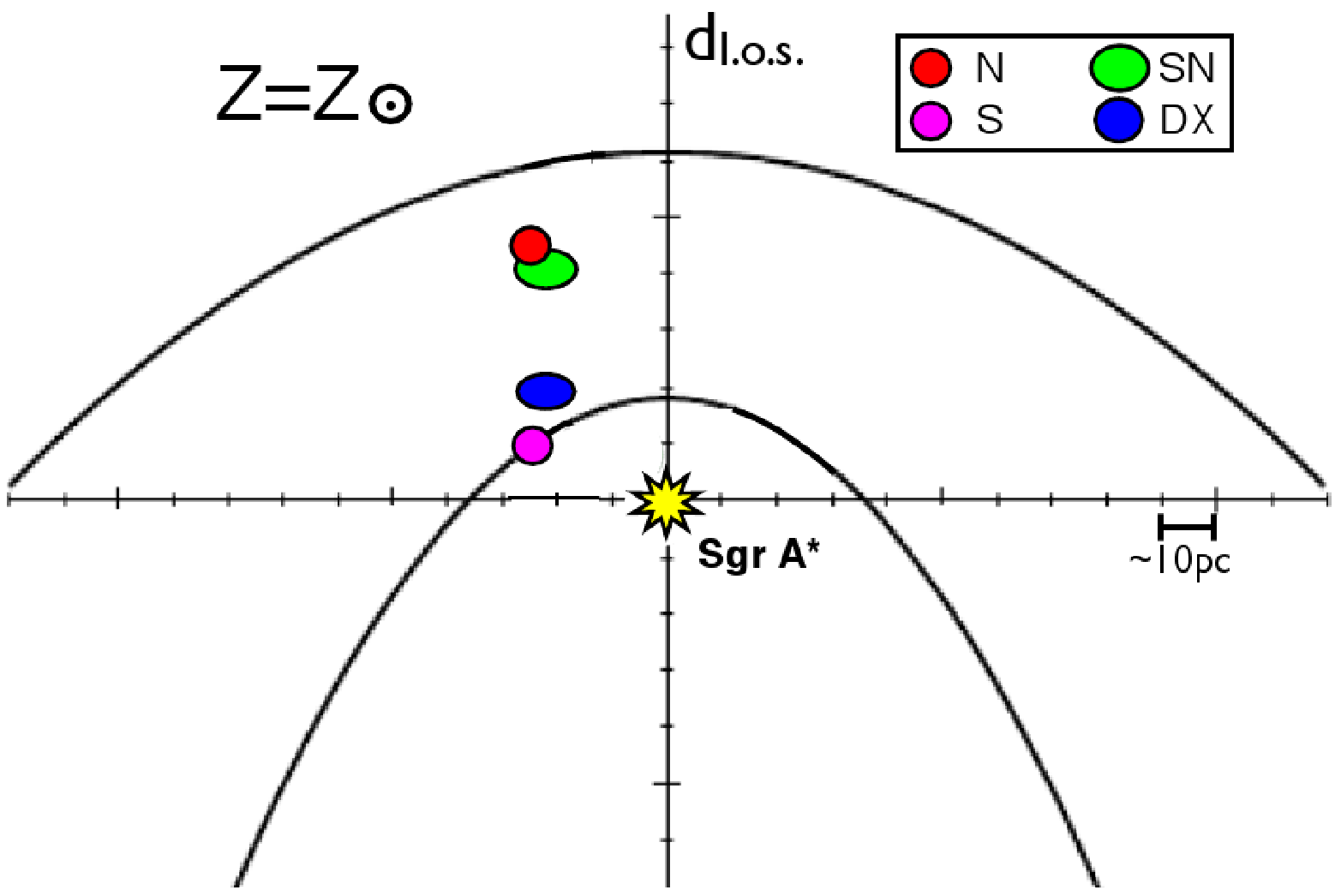} \hspace{1cm}
\includegraphics[width=0.4\textwidth]{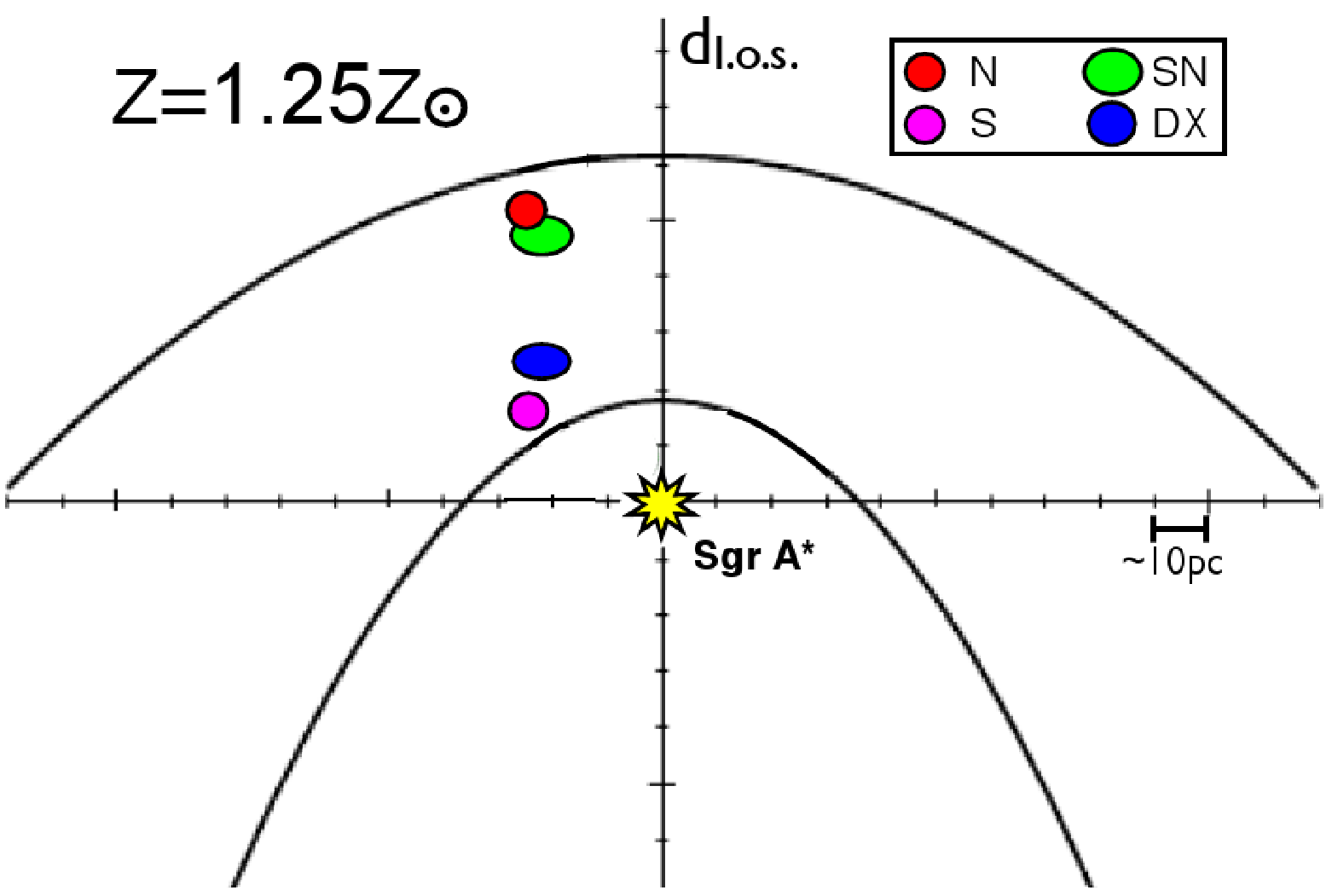}\\
\vspace{0.5cm}
\hspace{-0.09cm}
\includegraphics[width=0.4\textwidth]{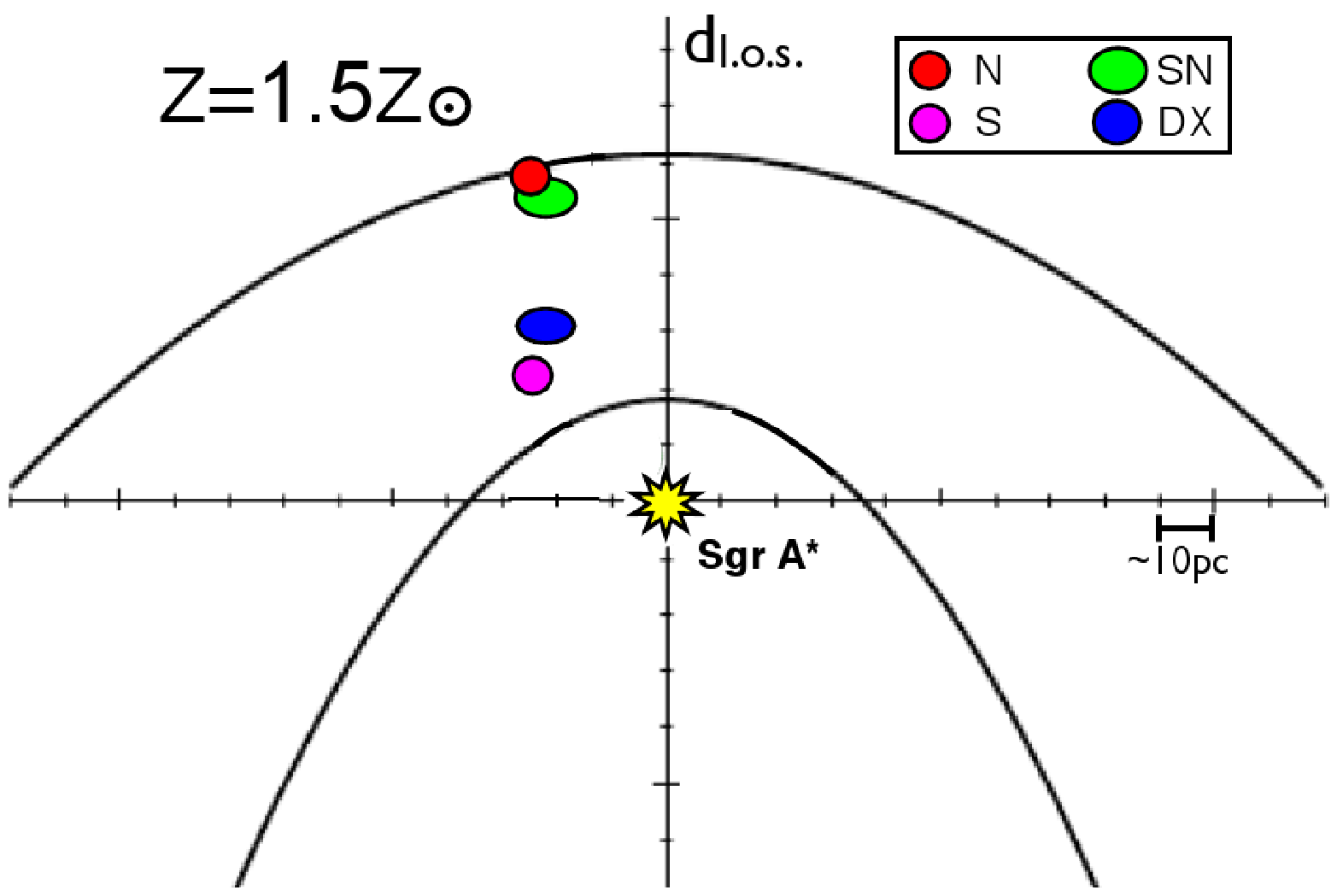}\hspace{1.12cm}
\includegraphics[width=0.4\textwidth]{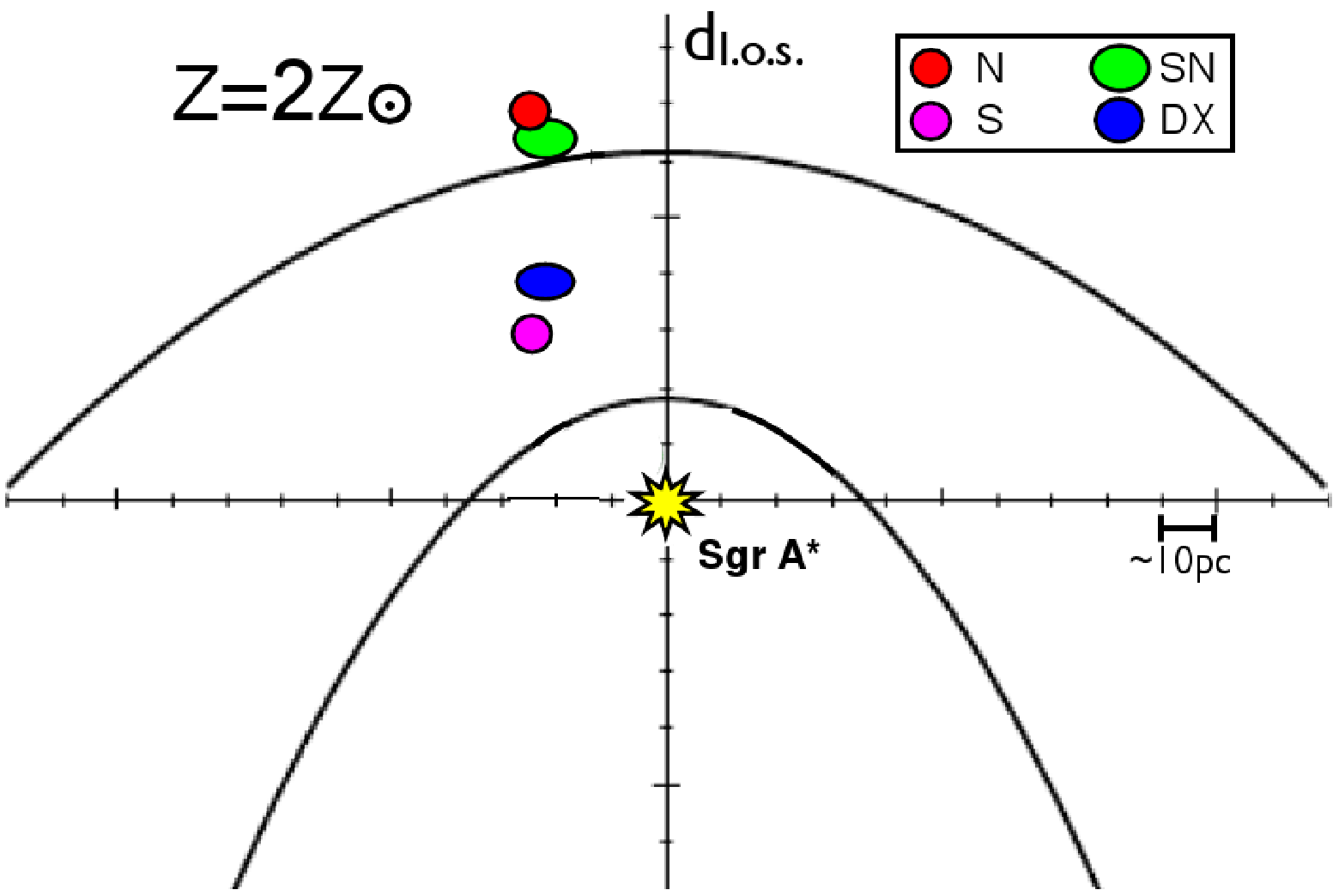}
\end{center}
\caption{The relative locations of the four MCs as inferred from the
XRN/Sgr A* outburst model suggested by \citet[][]{2010ApJ...714..732P}.
This is the view of the Galactic plane as seen from above (i.e.
the clouds positions are shown projected onto the plane).
The two parabolas in each 
sketch represent curves of constant path length for a burst of ionizing flux propagating outwards 
from Sgr A*.  The size of each cloud has been enlarged by a factor of 3.5 to aid the visualisation.
The distance (in parsecs) along the line of sight (relative to  Sgr A*) is plotted along the Y axis 
(see also Table \ref{clouds_distances}). The SMBH Sgr A* is located at the origin of the axes
and marked by a yellow star. The four panels represent the situation for an assumed metallicity of 1, 1.25, 1.5 and 2 times the solar value.}
\label{parabole}
\end{figure*}

In the XRN scenario, it has been conjectured that the most plausible
source of hard X-ray photons is Sgr A*.
On the basis of this assumption, we have 
investigated the possible distribution along the line of sight of the 6.4-keV
clouds in the AC vicinity. In the following calculations,  we assumed the 
X-ray luminosity of the   Sgr   A*   giant   flare  to have been  
1.4$\times$10$^{39}$   erg/s
\citep[][]{2010ApJ...714..732P}.   We then  infer  the   physical separation
between  the  clouds  and  Sgr  A*  using  the  formula  reported  in
\citet[][]{1998MNRAS.297.1279S} which, to better fit our purposes, can
be written in the form

\[ \mathrm{\frac{R^{2}}{4 d^{2}}=\Omega=\frac{4\pi D^{2}\cdot F_{6.4}}{\tau \cdot L_{x} \cdot 10^{7} \cdot Z} \hspace{0.1cm}= \hspace{0.1cm} 5.17\times 10^{-4}\left(\frac{F_{6.4}}{10^{-4}}\right) \left(\frac{0.1}{\tau}\right) \left(\frac{Z_{\odot}}{Z}\right). } \]

In this  equation  D is  the distance to 
the GC (assumed to  be 8 kpc), F$_{6.4}$
is the  Fe-K$_{\alpha}$ line flux (in  photons/cm$^{2}$/s), $\tau$ the
optical  depth to  Thomson scattering  of the  cloud, and  L$_{x}$ the
X-ray  luminosity   of  the   Sgr  A*  flare.  The solid  angle $\Omega$  is  defined as
$\Omega$=R$^{2}$/4d$^{2}$, where R is the  radius of the cloud as seen from Sgr A* (basically, the minor axis for the cloud DX and a combination of major and minor axis for the region SN, as measured on the sky) and d
is the distance of  the  cloud from Sgr A*
($\Omega$ is scaled to be the fraction of the whole sky 
subtended by the MC from the perspective of Sgr A*).

By first determining $\Omega$, we are then able to estimate d 
from the cloud dimension (assuming spherical symmetry).
By comparison of d with the projected distance of the cloud from Sgr A*
on the plane of the sky (d$_p$), we can then derive the position of the clouds
along the line of sight relative to Sgr A*  (d$_{l.o.s}$). 

Our calculations are, of course, highly dependent on the assumed
past lumininosity of Sgr A*. However, since we use the best estimate
available based on the XRN modelling of other GC clouds
\citep[][]{2010ApJ...714..732P}, we are in effect testing this model for
internal self-consistency. However, the inferred line of sight location of the clouds
(d$_{l.o.s}$) will depend on the assumed metallicity (Z) of the molecular 
material (more specifically on the iron abundance) and also on
the  optical depths of the MCs. In the literature, there is  not 
yet a consensus  on the typical values of these parameters for MCs in the CMZ.
In the case of the metal abundance, the best estimate is for
supersolar values,  closer to Z=2 than Z=1
\citep[][]{1996ARA&A..34..645M}.

All  the assumptions and  the results of our calculations
are   detailed  in   Table   \ref{clouds_distances}. The Fe-K$_{\alpha}$ line fluxes for the different clouds are assumed as in the weighted mean column of Table \ref{ka_fluxes}; because of the measured variability, only for the DX cloud we assumed the peak flux measured in September 2004 (see Table \ref{ka_fluxes}) of 4.7$\times$10$^{-6}$photons/cm$^{2}$/s. The inferred geometry
of the clouds distribution  is shown   in Fig.\ref{parabole}. Here, the cloud locations
relative to Sgr A* are shown projected onto the Galactic plane with the y-axis 
representing the direction along the line of sight. The two  parabolas 
correspond to  curves of constant light path, representative of the leading and trail
edge of the inferred outburst on Sgr A*, as discussed by \citet[][]{2010ApJ...714..732P}. The panels in Fig.\ref{parabole} illustrate the impact of different
assumptions with respect to the metallicity of the MCs.

We first consider the three clouds close to the AC (in projection), namely the N, S
and SN clouds.
Assuming a solar metallicity for the MCs, we infer line of sight displacements relative to Sgr A* 
(d$_{l.o.s}$) of 45, 9.9 and 41.4 pc for the N, S and SN knots (Fig.\ref{parabole}, 
upper left panel). In this scenario we have the N and SN clouds placed in the region between 
the two parabolas and therefore currently illuminated by the Sgr A* outburst. A relatively 
constant 6.4-keV line flux is consistent with this model (assuming that Sgr A* remained 
in a relatively constant high state for the duration of the outburst).
On the other hand, the cloud S is being crossed by the trailing edge of the irradiation burst. 
Therefore, one might expect that its Fe-K$_{\alpha}$ line flux 
to exhibit similar behaviour to that seen in the Sgr B2 and G0.11-0.11 MCs 
\citep[][respectively]{2009PASJ...61S.241I,2010ApJ...714..732P}, i.e., 
a steady decrease with time. Since such a decrease is not observed,  there is 
clearly some inconsistency in the model based on the solar metallicity assumption. 
Moreover, we notice that the molecular complexes N and S are probably
dense regions within the same larger MC \citep[the -30 km/s MC, see Fig.14 
in][]{2006MNRAS.371...38W}: therefore, the relative separation of these 
two complexes by about 30-40 pc along the line of sight seems unlikely.

We   therefore  repeated  the   calculations  assuming   different  Fe
abundances;  the  values we  considered  are  1.25,  1.5 and  2  times
solar ($d$ $\propto$ Z$^{0.5}$). The  results  
of these  calculations  are  detailed  in  Table
\ref{clouds_distances}, and illustrated in Fig.\ref{parabole}. We find
that  for  a metallicity  higher  than 1.5,  the  XRN  model again has
problems in explaining the constancy of the Fe-K$_{\alpha}$ lightcurves
of all the clouds. In the case of Z=1.5, while  the clouds  S and  SN lay  
well within the outburst region, the  N cloud is at its leading edge 
(lower left panel), implying an increasing
fluorescence signal \citep[a  behavior  similar  to the  one  measured from  the
\textit{bridge} in][]{2010ApJ...714..732P}. When Z=2 (lower right
panel) although the S region lies within the outburst region, 
the incident ionizing front propagating from Sgr A* has not yet reached the
N and SN knots.

For Z=1.25, the N, S and SN  clouds all lie in the outburst region. In
this case, and more generally for a metallicity range of 1.2-1.4 times
solar, the  XRN/Sgr A* outburst  model can, in principle,  explain the
observed fluorescence both in terms  of the observed line flux and the
lack of  strong variability. Within this framework,  the separation of
the N and S  clouds is inferred to be between 30-40  pc which seems to
be a  highly unlikely scenario (as noted  previously).  However, given
the  uncertainty in  measuring the  optical depths  of  the individual
clouds based on the available N$_{H}$ measurements, it  is at least
plausible that this unlikely cloud separation is a consequence of the
limited precision of the optical depth estimates.

An alternative  to the Sgr  A* outburst
hypothesis is that the energising photons are supplied by other nearby
sources; indeed  the GC region hosts numerous  X-ray transient sources
\citep[][]{2005ApJ...622L.113M,2006ApJS..165..173M,2005A&A...443..571P,2005A&A...430L...9P}.
We note that \citet[][]{2003ANS...324..117K} have studied the  diffuse
Fe-K features from the MCs in
the  AC surroundings using  Chandra which,  through its  sharp angular
resolution,  is better able  to delineate  the shape  of the  6.4- keV
bright regions.  These authors first suggested that  the line emission
is probably  due to  sources other  than a past  intense flare  on Sgr
A*. Particularly for the N and S MCs, the different surface brightness
of two  different regions within the  same MC \citep[the -30 km/s  MC,][]{2006MNRAS.371...38W}
would need to be explained in terms of  varying optical
depths if the  irradiation is from a single (relatively distant)
bright source. Given  the proximity of  these clouds to
the  AC, the  alternative  scenario of  bombardment  by low-energy  CR
electrons certainly merits consideration (see below).

\subsubsection{The peculiar case of the DX cloud} 
We now consider the DX cloud, which is \textit{unique} amongst the GC
MC that are 6.4-keV bright in that it has exhibited both an increase  and 
decrease in its Fe-K$_{\alpha}$ flux.  
The variability measured in the DX MC is too fast to have been produced
by  the same  Sgr A* flare which has been proposed to have illuminated 
the other XRN in the CMZ, since the inferred flare duration for the latter
is greater than ten  years.  Notwithstanding this constraint, we have determined 
the location along line  of sight of  the DX cloud
using the same  method as employed above  (see  Table
\ref{clouds_distances} for the detailed results).  
As we  can  see in  all  the panels  of Fig.\ref{parabole},  the
d$_{l.o.s}$ value for this cloud spans the range 20-40 pc, putting it
squarely within the outburst region for any of the metallicity assumptions.
Clearly, a simple XRN/Sgr A* picture as the one proposed by \citet[][]{2010ApJ...714..732P} is not consistent with our results on the DX cloud; on the other hand, we note that there is a source of potential errors introduced by the uncertain measurement of the distance along the line of sight. However, the fast variability measured for the Fe-K$_{\alpha}$ line flux points to an origin other than Sgr A*, whose flare luminosity is supposed to be constant over the whole flaring period. In this cloud, the non detection of the Fe-K edge is due to the poor statistics.

In the XRN hypothesis, the energising source could also be a transient
source, most likely an X-ray binary (XRB) system. The typical 2-10 keV
luminosities reached by these X-ray  point sources are of the order of
10$^{36}$ and  10$^{38}$ erg/s for a  high-mass and a  low-mass system
respectively. Assuming an XRB embedded in the DX cloud
($\Omega$=1 in the formula above), the required X-ray luminosity is  
$\sim$2$\times$10$^{35}$erg/s,  compatible with a transient
source hypothesis.  On  the other hand, if  we consider the possibility that the
primary  source might have been a low-mass XRB lying in a region of high obscuration
immediately behind the cloud, then it could be up to $\sim$10 pc from the cloud 
assuming a typical X-ray  luminosity of  10$^{38}$ erg/s.
Given the X-ray luminosity requirements, we favor the
scenario where a high-mass XRB is located inside, or closely behind,
the cloud.  Among such systems,  transients sources can  have orbital
periods   of  years,   with   a  flare   timescale   of  some   months.  
Of course, given the very limited information we have to date, 
we  cannot exclude the contributions of more than one X-ray source or even
perhaps a more exotic contribution to the fluorescence excitation from a short-lived 
particle bombardment episode.

\subsubsection{EW of the Fe-K$_{\alpha}$ line}
The relatively  high values of the  line EW measured in  the three MCs
closest to the AC are readily explained in terms of photoionization of
the clouds  by X-ray  photons. In the context of the XRN model, the EW of the Fe-K$_{\alpha}$ line is expected to be high (about 1 keV), since the direct source of photoionization 
is not seen by the observer \cite[e.g.,][]{1998MNRAS.297.1279S}. Therefore, our results seem to confirm that some kind of X-ray reflection is working in the MCs close to the AC; however, the lack of a strong present-day X-ray source leaves open the discussion about the identity of the putative energizing source.  Although the EW of the 6.4-keV line measured in the three MCs is about 1 keV, we notice that the error ranges (90\% confidence level) are always consistent with an EW value of 0.6-0.7 keV; this is what expected from fluorescence to be induced by particle bombardment (subrelativistic electrons) in a MC with an Fe abundance of about twice the solar value. We therefore cannot exclude a priori one of the two ionization mechanisms only based on the present measurements of the EW.

\subsubsection{Fe K absorption edge}
We did not measure any Fe-K absorption edge at 7.1 keV; although the detection of such a spectral feature is challenging for N$_{H}$ values of the clouds in the range 10$^{22}$-10$^{23}$cm$^{-2}$, we notice that in other GC clouds with similar density an Fe-K edge has been measured \cite[][]{2010ApJ...714..732P}. We could only find upper limits to the optical depth of the absorption edge, listed in Table \ref{number_sigma2}.

\subsubsection{Spectral hardness}
The shape of the non-thermal continuum associated with the
production of the 6.4-keV line can also, in principle,
provide a clue as to the nature of the excitation mechanism.
Our analysis suggests that this non-thermal continuum
has a very hard spectrum with a power-law photon index
$\Gamma$=1.0$^{+0.1}_{-0.2}$. 
In the particle bombardment scenario,
$\Gamma$=1.3-1.4 is expected for CR electron spectra
typical of the GC region \citep[e.g.][]{2002ApJ...568L.121Y},
although a somewhat harder spectrum might be indicative of
an unusual particle environment in the vicinity of the AC.
Therefore, the hardness of the power law component
might be best explained 
in terms of the bombardment of the clouds by cosmic ray
particles emanating from the AC itself.
Of course, the pure reflection scenario also
predicts a reflected X-ray continuum significantly
harder than that characteristic of the illuminating
X-ray source \citep[][]{2004A&A...425L..49R, 2010ApJ...719..143T}.

\subsection{The CR particle bombardment hypothesis}

An  alternative explanation of the Fe fluorescence 
involves the  interaction of CR particles  (electrons and/or protons)
with cold molecular matter. In this context the cross-section for collisional  
ionization of Fe has its  highest value in the  energy range extending
from the energy of the Fe-K edge (at 7.1 keV) up to a few MeV. Recently, 
two main  ideas have been developed in order to explain the Fe fluorescent 
emission, namely bombardment by either  electrons
\citep[][]{2007ApJ...656..847Y} or protons
\citep[][]{2009PASJ...61..901D}.  \citet[][]{2000ApJ...543..733V} first
suggested that a distribution of low-energy CR electrons 
(LECRe) might be responsible for the Galactic Ridge X-ray emission.
In their model, the hard X-ray continuum associated with the Ridge
is produced by non-thermal  bremsstrahlung as the LECRe interact with
the material of the ISM.  The low-ionization line emission 
from Fe and other metals is similarly produced through LECRe collisions 
with the neutral medium. The energies of the LECRe involved in this mechanism 
span the range 7.1 keV to $\sim$100 keV. In this setting, the flux of the 
6.4-keV line is directly proportional to the density of the target material
(in our case the MCs), and the in situ kinetic energy density of the LECRe, while the resulting EW of the 
Fe-K$_{\alpha}$ line spans the range 0.3-0.7 keV for metallicities from solar to twice the solar value.

The 6.4-keV photon  production rate inferred  from the 
results of \citet[][]{2000ApJ...543..733V} is

\[ \mathrm{F_{6.4} =\frac{4.8\times 10^{-22}H}{4\pi D^{2}}\left( \frac{U}{eV/cm^{3}}\right)  \hspace{0.2cm}photons/cm^{2}/s},
\]

\noindent  where  F$_{6.4}$ is  the  6.4-keV  line  flux  in units  of
photons/cm$^{2}$/s, D  is the distance to emitting cloud (8 kpc for the GC
clouds), U is the kinetic energy density of  LECRe (in eV/cm$^{3}$)
and H is the  number of H atoms in the MC  (see fourth column of Table \ref{clouds}).

Using the estimates of the number of H atoms given in Table \ref{clouds},
we have calculated  the LECRe energy density required to explain the observed
6.4-keV fluxes (Table \ref{ka_fluxes}).
We find that the energy density required in the  N, S  and SN  knots
is 61, 216  and 77 eV/cm$^{3}$.  For the DX  cloud, assuming  the Fe-line 
flux  to be 1.6$\times$10$^{-6}$photons/cm$^{2}$/s (i.e. the minimum in the lightcurve), 
the required energy density is 32 eV/cm$^{3}$. All these four values have to be divided by two in case the Fe abundance we assume is twice the solar value (as discussed above).
For comparison, \citet[][]{2000ApJ...543..733V}  found that
the mean LECRe energy  density required to explain the Galactic Ridge
was 0.2 eV/cm$^{3}$; moreover, using the same model, \citet[][]{2002ApJ...568L.121Y} found that the Fe-K$_{\alpha}$ line flux measured from the G0.13-0.13 MC requires a LECRe energy density of 150 eV/cm$^{3}$, a value very similar to what found for the AC MCs.

A significant enhancement of the CR flux in the  GC region compared to
the Galactic plane is perhaps to be expected. Also energetic particles 
might well be accelerated in the shocked  stellar winds which characterise 
the core region of the AC  \citep[][]{1985ApJ...289..698W,2003ApJ...590L.103Y}, potentially resulting 
in a greatly enhanced energy density in the vicinity of the cluster. 
The fact that the 6.4-keV line flux measured  from region S is significantly 
higher than that from knot N could be  a consequence  of the
stellar   distribution  within   the  cluster.   As  shown   first  by
\citet[][]{2002ApJ...570..665Y}          and          later         by
\citet[][]{2004ApJ...611..858L}, the southern component of the cluster
is  populated by strong stellar-wind  sources, which are also detectable
at   radio  wavelengths  \citep[][]{2001ApJ...551L.143L}.  These
massive early-type stars  could also be present in  binary systems, and
be the origin of the X-ray  flaring  activity recently
discovered  by  \citet[][]{2011A&A...525L...2C}.  Massive  interacting
binaries are efficient in producing X-ray emission in the shock fronts
within the  winds and also in accelerating CRs.

In a variation on this theme, \citet[][]{2006MNRAS.371...38W}  proposed the interaction
of the -30 km/s MC  with the AC to be the origin  of the Fe fluorescent
emission. These  authors found that  the -30km/s molecular  complex is
moving towards the  AC with a relative velocity  of $\sim$120 km/s. In
this  picture, particles  are accelerated  in reversed  shocks through
efficient Fermi  first order mechanism,  and gain enough energy  to be
able to ionize Fe atoms within the cloud.

In the case of the  DX cloud, we  discovered the $\textit{fastest}$ variability
ever found  for a GC MC. Given that the timescale of the variability is
comparable to the size of the cloud, this would seem  to completely rule  out
bombardment by sub-relativistic particles as the origin of  the increase  in the  
Fe-K$_{\alpha}$ line flux  measured  from this  complex. 
However, it is at least  plausible that the underlying base level of the 6.4-keV 
lightcurve could represent a relatively constant particle contribution to
the fluorescence budget. The presence of high energy particles 
in  this cloud may be inferred from radio continuum observations at 20  cm \citep[see Fig.1b
in][]{2002ApJ...570..665Y} in which non-thermal  filaments propagate
westwards from  the Radio  Arc into the general region of this cloud.

The  nature  of  the  non-thermal  radio
filaments  in  the  GC   region  is  still  under  debate.  The
concentration of  these strong radio  features in the inner  Galaxy may well
be related to  the peculiar high energy activity of this
unique region \citep[][]{2003ApJ...598..325Y}. Many attempts have been made
to  explain the  formation of  such structures, involving all the likely high 
energy sources connected with the GC region, that is the SMBH Sgr A*, massive wind 
binaries, Galactic winds and gas clouds. A recent  suggestion is that the origin of 
these filaments might be due to the presence of many young and massive stellar clusters
in the  central  region  of  the  Galaxy \citep[][]{2003ApJ...598..325Y}. Whatever
the origin of the radio filaments, their presence in the region suggests
that cloud fluorescence induced by particle bombardment must be present at some
level \citep[i.e.][]{2009PASJ...61..593F}.
Potentially, X-ray flaring events of the sort recently
discovered  in   the  AC  \citep[][]{2011A&A...525L...2C}, might also
induce some variablity in the local LECRE energy density, which could in turn impart 
variability in the fluorescence signal, albeit  delayed and smeared out
relative to the activity in the primary source.

Recently, \citet[][]{2009PASJ...61..901D} have suggested that at least some of the Fe-K$_{\alpha}$ line
emission from MCs in the CMZ can be produced by the interaction of high energy protons (with energies in
the range of several hundred MeV) with the molecular target. This model naturally explains the presence 
of both the non-thermal components in the spectrum of the GC hard X-ray emission \citep[][]{2009PASJ...61S.255K}. 
As in the LECRe model, this scenario has some difficulties in accounting for the rapid variability of the 
6.4-keV line found in some MCs in the inner GC region, including the DX cloud.

LECRe with E$\sim$100 keV can be stopped by matter with column densities of about 10$^{22}$ H atoms/cm$^{2}$ 
\citep[][]{2003EAS.....7...79T}. Considering the sizes and column densities of the MCs quoted in 
Table \ref{clouds}, we can calculate that LECRe loose their energy within depths that are at most 
half the cloud size. Similarly, \citep[][]{2009PASJ...61..901D} estimate that high energy protons 
in the GC region typically travel for 0.1-0.3 pc before being stopped through interactions with the ISM.  
These estimates imply that particle-induced  fluorescence and other interactions will occur 
only in the outer shell of a dense cloud.
Both LECRe and high energy protons lose most of their energy in heating the ISM rather than in producing 
non-thermal continuum emission or fluorescence photons.
\citet[][]{2007ApJ...665L.123Y} argue that the reason the temperature of the MCs in the CMZ is higher 
(T$\sim$100-200 K) than that measured for MCs in the Galactic plane (about 20 K)
is due to CR particle heating.  In fact, the ionization rates modelled for LECRe and high energy proton
bombardment of MCs are in good agreement with the temperature enhancements which are actually measured 
\citep[][]{2002ApJ...568L.121Y,2009PASJ...61..901D}.  Furthermore, both 
\citet[][]{1993A&A...280..255H} and \citet[][]{2001A&A...365..174R} measured 
temperature gradients in most of the MCs in the GC region; this is a natural consequence 
of the interaction of subrelativistic electrons and protons largely with the 
surface layers of the MCs.

\section{Summary}

In  this  paper we have used data from XMM-Newton to study  the  diffuse Fe-K$_{\alpha}$  
fluorescent line emission emanating from the region surrounding the  AC. 
Our results can be summarized as follows:

\begin{itemize}

\item{We detected four Fe-K$_{\alpha}$ bright MCs  in the  vicinity of the AC.
Two of these are new detections (referred to herein as knots SN and DX).}

\item{The Fe-K$_{\alpha}$ line  flux measured from the three  clouds nearest to the
AC (knots N, S and SN) has remained constant over an interval of eight
years. The EW of the 6.4-keV  line in the spectra of these clouds is
measured to be about 1 keV.  This is consistent with the origin of the
Fe  fluorescence being  the irradiation  of  the MCs  by X-rays.   The
XRN/Sgr A* outburst scenario, which has been recently suggested as the
explanation for  all the  Fe-K fluorescence seen  in the GC  region, 
in broad terms fits the observations,   although   with  some
reservations.   Other  X-ray   sources,  in  particular  nearby  X-ray
binaries and  transients, might well  have contributed to the  6.4 keV
line  production.  Within the  errors,  the  Fe-K$_{\alpha}$  line EWs  are  also
compatible with the CR bombardment  scenario, with the AC itself being
a  likely location of  the requisite  particle acceleration.
In this context, the required particle kinetic
energy density is roughly a hundred times higher than that previously estimated
as a possible explanation of the hard X-ray Galactic Ridge emission, but
such an enhancement might well arise from in situ acceleration of particles 
in the AC. The particle candidates are subrelativisitc electrons and/or protons.}  

\item{We have discovered variability on a timescale of a few years
in the Fe-K$_{\alpha}$ line emission from the DX cloud. Because of the highly variable emission, this is unlikely to 
be the product of  past high-state activity in Sgr A*. We propose an 
X-ray binary to   be   the   energising  source   of   this   fluorescence.
Moreover, on  the  top of  the  variability, this  molecular
complex  shows  a  non-zero underlying level of the fluorescent line  flux,
suggesting the possibility that both the reflection and CR bombardment 
processes may be working in tandem. The variability seen in the DX cloud 
is the $\textit{fastest}$ yet recorded in the GC region and, to date, it is
the  only example of a MC exhibiting  both an increase and subsequent
decrease in its Fe fluorescent emission.}

\end{itemize}

The  available  data for the AC region does not allow the
precise measurement of the spectral   parameters 
relating to the Fe fluorescence phenomenon, namely the EW of the Fe K$_{\alpha}$ line and the absorption edge at 7.1 keV.  However, with deeper
observations there is the prospect of fully characterising the
emission spectrum and dispelling any ambiguity concerning
the underlying excitation process. This would be an important
further step in the on-going investigation of whether Sgr A* 
has exhibited AGN-like activity in the recent past.

\begin{acknowledgements}
XMM-Newton is an ESA science mission with instruments and contributions directly funded by ESA Member States and the USA (NASA).
R.C. thanks Prof. Dr. Y. Tanaka and Dr. Lara Sidoli for many useful discussions. R.C. also thanks Dr. Konrad Dennerl and Dr. Silvano Molendi for useful discussions about the XMM-Newton background subtraction and modelling. 
\end{acknowledgements}



\end{document}